\documentclass[12pt]{article}
\usepackage{latexsym,graphicx,epsfig,psfrag,here}
\newcommand{\beq}{\begin{equation}} 
\newcommand{\eeq}{\end{equation}} 
\newcommand{\beqa}{\begin{eqnarray}} 
\newcommand{\eeqa}{\end{eqnarray}} 
\newcommand{\beqan}{\begin{eqnarray*}} 
\newcommand{\eeqan}{\end{eqnarray*}} 
\newcommand{\ba}{\begin{array}} 
\newcommand{\ea}{\end{array}} 
\newcommand{\no}{\nonumber} 

\newcommand{\lets}{\stackrel{<}{_\sim}}
\newcommand{\Un}{\underline} 
\newcommand{\ol}{\overline}

\newcommand{\ve}{\varepsilon}

\newcommand{\wt}{\widetilde} 
\newcommand{\wh}{\widehat}

\newcommand{\cL}{{\cal L}} 
\newcommand{\M}{{\cal M}}

\newcommand{\mbf}{\mathbf} 
 
\newcommand{\dfrac}{\displaystyle \frac} 
 
\newcommand{\dsum}{\displaystyle \sum} 
\newcommand{\dprod}{\displaystyle \prod}

\newcommand{\nn}{\nonumber \\}

\newcommand{\bea}{\begin{eqnarray}} 
\newcommand{\eea}{\end{eqnarray}}

\newcommand{\hepph}[1]{{\tt hep-ph/#1}} 
 
\newcommand{\PL}[3]{{Phys. Lett.} {\bf#1,} {#2} {(#3)}} 
\newcommand{\PRL}[3]{{Phys. Rev. Lett.}  {\bf#1,} {#2} {(#3)}} 
\newcommand{\PR}[3]{{Phys. Rev.} {\bf#1,} {#2} {(#3)}} 
\newcommand{\NP}[3]{{Nucl. Phys.} {\bf#1,} {#2} {(#3)}} 
\newcommand{\EPJ}[3]{{Eur. Phys. J.} {\bf#1,} {#2} {(#3)}} 
\newcommand{\ZP}[3]{{Z. Phys.} {\bf#1,} {#2} {(#3)}} 
\normalsize 
\begin{document} 
\begin{titlepage} 
\begin{flushright} 
UWThPh-2001-55\\ 
Feb. 2002\\ 
\end{flushright} 
\vspace{2.5cm} 
\begin{center} 
{\Large \bf Four-Pion Production in $\mbf{e^+e^-}$ 
Annihilation*} \\[40pt] 
G. Ecker and R. Unterdorfer  
 
\vspace{1cm} 
Institut f\"ur Theoretische Physik, Universit\"at 
Wien\\ Boltzmanngasse 5, A-1090 Vienna, Austria \\[10pt] 
 
\vfill 
{\bf Abstract} \\ 
\end{center} 
\noindent
We calculate the processes $e^+ e^- \to  4\pi$ and 
$\tau \to \nu_\tau 4 \pi$ 
to $O(p^4)$ in the low-energy expansion of the standard
model. The chiral amplitudes of $O(p^4)$ can be extended via
resonance exchange to energies around 1 GeV. Higher-order effects
have been included in the form of $\omega$, $a_1$ and double 
$\rho$ exchange and by performing a resummation of the pion form
factor. The predicted cross sections and the branching ratios 
$BR(\rho^0 \to 4 \pi)$ are in good agreement with available data. 
 
\vfill 
\noindent *~Work supported in part by TMR, EC-Contract  
No. ERBFMRX-CT980169 (EURODA$\Phi$NE) and by a Research Fellowship of
the Univ. of Vienna.
 
\end{titlepage}
\section{Introduction}
\label{sec:Intro}
\renewcommand{\theequation}{\arabic{section}.\arabic{equation}}
\setcounter{equation}{0}
Electron positron annihilation into hadrons has played an important
role in the development of modern particle physics. Precise knowledge
of the cross section is essential for many purposes, in
particular for the determination of the hadronic contribution to
the anomalous magnetic moment of the muon and for running the 
fine-structure constant up to $M_Z$ to analyse electroweak precision
measurements. 

At high energies, the inclusive cross section can be calculated in QCD.
It provides one of the standard tests of QCD allowing for the
extraction of
the strong coupling constant. At energies below approximately 2 GeV,
the different exclusive channels are measured separately. As
perturbative QCD cannot be applied to those exclusive processes the
theoretical challenge consists in modeling them in a way that is at
least consistent with QCD. 

At very low energies ($E <<$ 1 GeV), the
most reliable approach is furnished by chiral perturbation theory
(CHPT), the systematic low-energy effective theory
\cite{chpt,gl84,gl85a} of the standard model. 
Although the low-energy expansion of CHPT breaks down at typical
hadronic scales of $O(M_\rho)$ it can still provide important
constraints how to match the low-energy amplitudes on to the
intermediate-energy region governed by meson resonance exchange 
\cite{egpr89,eglpr89}. A simple but very illustrative example is the
pion form factor measured in $e^+ e^- \to \pi^+ \pi^-$ that can be
continued from threshold to the region beyond 1 GeV in a
straightforward way \cite{gp97,gdpp00,pff2}.

Inspired by this success and by the phenomenological importance of
four-pion production, especially
for the calculation of $\alpha(E)$, we have undertaken a systematic
study of $e^+ e^- \to 4 \pi$ in CHPT (with the two possible charge
configurations $2\pi^0\pi^+\pi^-$, $2\pi^+ 2\pi^-$). At first sight,
this does not appear very promising with the threshold 
$E=4 M_\pi\simeq 560$ MeV already in the vicinity of the most
prominent meson resonance, the $\rho$ meson. However, once again the
chiral amplitudes of $O(p^4)$ can be continued into the
resonance region and the decay rates $\Gamma(\rho\to 4 \pi)$ can be
calculated with reasonable accuracy. We will also consider the energy
dependence of cross sections for $E \le$ 1 GeV.

There is an essential difference between the two- and
four-pion modes. Whereas the two-pion amplitude is completely
dominated by $\rho$ exchange even beyond 1 GeV the situation is much
more involved for four-pion final states where at least
$\rho$, $\omega$ and $a_1$ exchange are relevant (e.g., 
Refs.~\cite{dhjf96,ck01,novosibirsk}). It is then all the more 
important to have
unambiguous theoretical guidelines for the construction of those
amplitudes such as the correct low-energy behaviour dictated by QCD.

In addition to electron positron annihilation, multi-pion final states
can also be studied in $\tau$ decays (for reviews of the theory see,
e.g., Refs.~\cite{pichrev,kuehnrev}). In the limit of isospin symmetry
that will be assumed throughout this paper 
both the two- and the four-pion modes are related. There is again an
important difference between the two modes. Whereas the annihilation
amplitude and the decay amplitude for two pions in the final states 
are in one-to-one correspondence the situation is more subtle in the
four-pion case \cite{kuehnrev}: given the amplitude for 
$e^+ e^- \to 2 \pi^0 \pi^+ \pi^-$ or for $\tau^- \to \nu_\tau 2 \pi^-
\pi^+ \pi^0$, the three remaining annihilation and decay amplitudes 
are uniquely determined but not vice versa. Therefore, in the
isospin limit it is sufficient for a complete
determination of all four amplitudes to construct the amplitude for the
$2 \pi^0 \pi^+ \pi^-$ channel only. All calculations in this paper
will be performed for this particular channel.

In Sec.~\ref{sec:kin} we collect the kinematics, matrix elements and
cross sections for the process $e^+ e^- \to 4 \pi$. We recall the
isospin relations relating $e^+ e^-$ annihilation and $\tau$ decays
into four pions. To make Bose symmetry and $C$ invariance manifest, we
express all matrix elements in terms of a reduced amplitude that
reduces the size of the various amplitudes roughly by a factor
four. The leading-order amplitudes of $O(p^2)$ for both 
$e^+ e^-$ annihilation and $\tau$ decays are presented in
Sec.~\ref{sec:p2}. The matrix elements of $O(p^4)$, consisting of both
one-loop and tree-level contributions, are calculated in 
Sec.~\ref{sec:p4}. The structure of the local amplitude of $O(p^4)$
determines the resonance exchange amplitudes generated by
$\rho$ and scalar exchange. The relevant chiral resonance Lagrangian
and the resulting matrix elements are presented in
Sec.~\ref{sec:reso}. To extend the amplitudes into the resonance
region, additional contributions are necessary. In
Sec.~\ref{sec:beyond} we analyse double $\rho$, $\omega$ and $a_1$ 
exchange to obtain the final amplitude. We collect the leading terms 
in the low-energy expansion of the pion form factor and we resum 
those terms to the complete $\rho$ dominated form factor. 
The energy dependent cross sections
for the two channels and the partial widths $\Gamma(\rho^0 \to 2 \pi^0
\pi^+ \pi^-)$, $\Gamma(\rho^0 \to 2 \pi^+ 2 \pi^-)$ are analysed in
Sec.~\ref{sec:rates}. We compare our results with available data in
the region 0.65 $\le E$(GeV) $\le$ 1.05 and we discuss the necessary
steps for proceeding to higher energies. Our conclusions are
summarized in Sec.~\ref{sec:conc}. Three appendices contain 
a discussion of the isospin relations, a brief
summary of the two possibilities for incorporating spin-1 mesons in 
chiral Lagrangians and a collection of numerical input for the 
calculation of cross sections.

\section{Kinematics and symmetries}
\label{sec:kin} 
\renewcommand{\theequation}{\arabic{section}.\arabic{equation}}
\setcounter{equation}{0}
The amplitude for the process
$$
e^+(k_+) e^-(k_-) \to \pi(p_1) \pi(p_2) \pi(p_3) \pi(p_4)
$$
is written in the form
\begin{equation} 
\M = \displaystyle\frac{e^2}{q^2 + i \ve}\ol v(k_+)\gamma_\mu
u(k_-) J^\mu(p_1,p_2,p_3,p_4) ~,
\hspace*{1cm} q=k_+ + k_- = \dsum_{i=1}^4 p_i~,
\end{equation} 
with $J^\mu$ the pionic matrix element of the electromagnetic current
\begin{equation} 
 J^\mu(p_1,p_2,p_3,p_4) = \langle \pi(p_1) \pi(p_2) \pi(p_3) \pi(p_4)
|J^\mu_{\rm elm}(0)|0\rangle ~.
\end{equation} 
The differential cross section is then given by (setting $m_e=0$)
\begin{equation} 
d\sigma = \displaystyle\frac{\alpha^2}{32 \pi^6 q^6}
(\dprod_{i=1}^4\frac{d^3 p_i}{2 E_i})\delta^{(4)}(q-\dsum_{i=1}^4 p_i)
l^{\mu\nu} J_\mu J_\nu^* \label{eq:csec}
\end{equation} 
with the leptonic tensor 
\begin{equation} 
l^{\mu\nu} = k_+^\mu k_-^\nu + k_-^\mu k_+^\nu - 
\frac{q^2}{2}g^{\mu\nu}~.
\end{equation} 
We will only be interested in the integrated cross sections 
$\sigma(q^2)$ where the appropriate statistical factors have to be
applied for the two possible channels $2\pi^0\pi^+\pi^-$ and
$2\pi^+ 2\pi^-$. 

With the charge assignments
$$
\pi^0(p_1) \pi^0(p_2) \pi^+(p_3) \pi^-(p_4)~, \hspace*{1cm} 
\pi^+(p_1) \pi^+(p_2) \pi^-(p_3) \pi^-(p_4)~,
$$
a convenient set of Dalitz variables\footnote{The main convenience is
in making symmetries manifest as discussed below.} is
\begin{eqnarray} 
q^2~, & s=(p_1 + p_2)^2 ~, & \nu=(p_3 - p_4)\cdot (p_1 - p_2)/2 ~, \nn
& t_i = p_i\cdot q ~~(i=1,\dots,4)  ~. & 
\end{eqnarray} 
There is a redundancy due to the relation $\displaystyle\sum_{i=1}^4
t_i = q^2$ but especially for displaying symmetries of the amplitudes
it is useful to keep the complete set. For compactness of
notation, we will often express amplitudes in terms of the various
scalar products instead of using $s$ and $\nu$.

In the isospin limit that will be assumed throughout this paper, the
current matrix element for the $2\pi^+ 2\pi^-$ channel can be
expressed in terms of the matrix element for the $2\pi^0 \pi^+ \pi^-$ 
channel
\cite{kuehnrev}:
\begin{eqnarray} 
\hspace*{-.5cm} \langle \pi^0(p_1) \pi^0(p_2) \pi^+(p_3) \pi^-(p_4)
|J^\mu_{\rm elm}(0)|0\rangle &:= & J^\mu(p_1,p_2,p_3,p_4) 
\label{eq:k1}\\
\hspace*{-.5cm} \langle \pi^+(p_1) \pi^+(p_2) \pi^-(p_3) \pi^-(p_4)
|J^\mu_{\rm elm}(0)|0\rangle &= & J^\mu(p_1,p_3,p_2,p_4) +
J^\mu(p_1,p_4,p_2,p_3) \nn
& +& J^\mu(p_2,p_3,p_1,p_4) + J^\mu(p_2,p_4,p_1,p_3)~. 
 \label{eq:k2}
\end{eqnarray}  
Likewise, the matrix elements of the charged vector current relevant
for $\tau$ decay can also be expressed in terms of 
$J^\mu(p_1,p_2,p_3,p_4)$ \cite{kuehnrev}:
\begin{eqnarray} 
\hspace*{-.7cm} \langle \pi^-(p_1) \pi^-(p_2) \pi^+(p_+) \pi^0(p_0)
|V^\mu_{\rm cc}(0)|0\rangle &= &\sqrt{2}\left\{J^\mu(p_+,p_1,p_2,p_0) 
+ J^\mu(p_+,p_2,p_1,p_0)\right\}  \label{eq:k3}\\
\hspace*{-.7cm} \langle \pi^0(p_1) \pi^0(p_2) \pi^0(p_3) \pi^-(p_-)
|V^\mu_{\rm cc}(0)|0\rangle&= &\sqrt{2}\left\{J^\mu(p_1,p_2,p_-,p_3) 
\right.  \nn
& +& \left. J^\mu(p_1,p_3,p_-,p_2) +  
 J^\mu(p_2,p_3,p_-,p_1)\right\}~,  \label{eq:k4}\
\end{eqnarray}
with the usual normalization of the charged vector current 
$V^\mu_{\rm cc}=\ol d \gamma^\mu u$ in terms of quark fields.
We come back to these matrix elements in Sec.~\ref{sec:p2} for 
the lowest-order chiral expansion.

In addition to isospin, the electromagnetic current matrix elements
are also constrained by gauge invariance, Bose symmetry and $C$ 
invariance. It is
sufficient to consider these symmetries in the 
$2 \pi^0 \pi^+ \pi^-$ channel. Via the isospin relation (\ref{eq:k2}),
the matrix element for the $2 \pi^+ 2 \pi^-$ final state is then 
automatically gauge invariant, Bose symmetric and odd under $C$. 
Of course, all these symmetries are implemented in CHPT so the
following relations emerge automatically in the calculation and 
need not be imposed a posteriori.

Gauge invariance (vector current conservation) implies
\begin{equation} 
q_\mu J^\mu(p_1,p_2,p_3,p_4) = 0 ~.\label{eq:vcc}
\end{equation} 
The remaining symmetry constraints for $J^\mu(p_1,p_2,p_3,p_4)$,
\begin{eqnarray} 
{\rm Bose~~symmetry:} & J^\mu(p_1,p_2,p_3,p_4)=J^\mu(p_2,p_1,p_3,p_4)
\\
C {\rm~~invariance:} & J^\mu(p_1,p_2,p_3,p_4)=-J^\mu(p_1,p_2,p_4,p_3)~,
\end{eqnarray} 
can always be made manifest by writing
\begin{eqnarray} 
J^\mu(p_1,p_2,p_3,p_4) &=&A^\mu(p_1,p_2,p_3,p_4)+
A^\mu(p_2,p_1,p_3,p_4) \nn
 && - A^\mu(p_1,p_2,p_4,p_3) -
A^\mu(p_2,p_1,p_4,p_3) \label{eq:Adef}
\end{eqnarray}
in terms of a reduced amplitude $A^\mu(p_1,p_2,p_3,p_4)$ that is 
not further constrained by $C$ invariance or Bose symmetry. In this
paper, we shall express all amplitudes in terms of 
$A^\mu(p_1,p_2,p_3,p_4)$. This makes the sometimes quite elaborate
matrix elements considerably more compact. Another simplification 
consists in dropping terms in $A^\mu(p_1,p_2,p_3,p_4)$ and therefore
also in $J^\mu(p_1,p_2,p_3,p_4)$ that are proportional to $q^\mu$. Of
course, such terms cannot contribute to the differential cross section
(\ref{eq:csec}). Dropping such terms may lead to seeming violations of
gauge invariance. It goes without saying that those terms can always
be recovered uniquely for a given matrix element by imposing current
conservation (\ref{eq:vcc}). This trivial remark will be relevant when
calculating $\tau$ decay matrix elements via the isospin relations
(\ref{eq:k3},\ref{eq:k4}).

\section{Amplitudes at leading order}
\label{sec:p2} 
\renewcommand{\theequation}{\arabic{section}.\arabic{equation}}
\setcounter{equation}{0}
\begin{figure}
\centerline{\epsfig{file=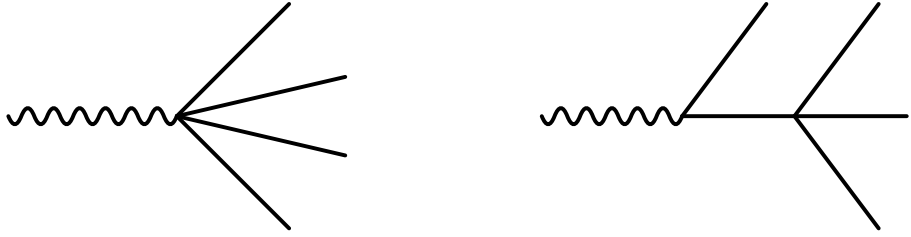,height=3cm}}
\caption{Tree diagrams for $\gamma^* \to 4 \pi$. In this and
subsequent figures solid lines denote pions; the wavy line stands for
a virtual photon.}
\label{fig:tree}
\end{figure}
At leading order in the low-energy expansion of the standard model,
$O(p^2)$,
the amplitudes for $e^+ e^- \to 4 \pi$ are determined by ``virtual''
bremsstrahlung. The corresponding diagrams shown in
Fig.~\ref{fig:tree} are easily calculated from the chiral Lagrangian
of $O(p^2)$ for chiral $SU(2)$ \cite{gl84}
\begin{equation} 
{\cal L}_2 = \frac{F^2}{4} \langle D_\mu U D^\mu U^\dagger +
             \chi U^\dagger + \chi^\dagger U \rangle ~.\label{eq:L2}
\end{equation} 
The notation is standard (see, e.g., Ref.~\cite{chptrev}).
For our purposes, the covariant derivative of the pion matrix field
$U$ contains only the external electromagnetic field $A^\mu$.  The
scalar field $\chi$ is proportional to the light quark mass $\hat m$
(we set $m_u=m_d:=\hat m$) and $\langle \dots \rangle$ 
denotes the 2-dimensional trace:
\begin{eqnarray} 
D^\mu U &=& \partial^\mu U + \frac{i}{2} e A^\mu [\tau_3, U] \nn
\chi &=& 2 B \hat m ~{\bf 1} = M^2  ~{\bf 1}=
M_\pi^2 [1 + O(\hat m)]  ~{\bf 1}  \nn
F_\pi &=&  F[1 + O(\hat m)] = 92.4 ~{\rm MeV} \nn
\langle 0|\bar u u |0\rangle &=& - F^2 B[1 + O(\hat m)] ~.
\end{eqnarray} 

As discussed in Sec.~\ref{sec:kin}, we express our results in terms of
the basic amplitude $A^\mu(p_1,p_2,p_3,p_4)$ defined in
(\ref{eq:Adef}) that determines all matrix elements of interest
(\ref{eq:k1}), \dots, (\ref{eq:k4}). For the tree-level amplitude
of $O(p^2)$ corresponding to the diagrams of Fig.~\ref{fig:tree}
one finds
\begin{equation} 
A^\mu_{(2)}(p_1,p_2,p_3,p_4)= \displaystyle\frac{s-M_\pi^2}
{F_\pi^2} \displaystyle\frac{p_3^\mu}{2 t_3 - q^2} ~.
\label{eq:Ap2}
\end{equation}
The complete current matrix element (\ref{eq:k1}) at $O(p^2)$ is 
therefore given by 
\begin{equation} 
J^\mu_{(2)}(p_1,p_2,p_3,p_4)= 
\displaystyle\frac{s-M_\pi^2}{F_\pi^2}
\left(\displaystyle\frac{2 p_3^\mu}{2 t_3 - q^2} -
\displaystyle\frac{2 p_4^\mu}{2 t_4 - q^2}\right)
~. \label{eq:Jp2}
\end{equation}  
We have used the physical quantities $M_\pi, F_\pi$ in these matrix
elements. The renormalization of $M,F$ to $M_\pi,F_\pi$ is an effect
of at least $O(p^4)$ and will of course be included in the amplitudes
of next-to-leading order.

The matrix element (\ref{eq:Jp2}) has an obvious interpretation:
$(s-M_\pi^2)/F_\pi^2$ is the leading-order amplitude for $\pi^0 \pi^0
\to \pi^+ \pi^-$ and the second factor reduces to the usual
bremsstrahlung factor for real photons ($q^2 \to 0$). Although we do
not discuss $\tau \to 4 \pi$ decays in any detail here, we also
display the tree-level current matrix elements (\ref{eq:k3}),
(\ref{eq:k4}), after repairing gauge invariance in (\ref{eq:Jp2}) by
adding the appropriate amplitude proportional to $q^\mu$:
\begin{eqnarray} 
\langle \pi^-(p_1) \pi^-(p_2) \pi^+(p_+) \pi^0(p_0)
|V^\mu_{\rm cc}(0)|0\rangle &=&  \nn
&& \hspace*{-7.5cm} \displaystyle\frac{\sqrt{2}}{F_\pi^2}\left\{- 2 (
p_+^\mu+p_0^\mu) + 2 R^\mu(p_0) p_+\cdot(q-p_0)
+\displaystyle\sum_{i=1}^{2}R^\mu(p_i)
[2 p_0\cdot (q-p_i)- M_\pi^2]\right\}  \label{eq:tau1} \\
\langle \pi^0(p_1) \pi^0(p_2) \pi^0(p_3) \pi^-(p_-)
|V^\mu_{\rm cc}(0)|0\rangle  &=&  \nn
&& \hspace*{-6cm} \displaystyle\frac{\sqrt{2}}{F_\pi^2}\left\{4 p_-^\mu-
M_\pi^2 R^\mu(p_-) - \displaystyle\sum_{i=1}^{3}R^\mu(p_i)
[2 p_-\cdot (q-p_i) - M_\pi^2]\right\} \label{eq:tau2} \\
R^\mu(p) &= & \displaystyle\frac{q^\mu - 2 p^\mu}{q^2 - 2 p\cdot q} 
~.\no 
\end{eqnarray} 

In the chiral limit ($M_\pi=0$), these matrix elements have the same 
structure as in the standard reference on the subject \cite{fww80}.
There are two misprints in Ref.~\cite{fww80} that have
propagated into some of the subsequent literature: 
Eqs.~(2.9a) and (2.9b) of Ref.~\cite{fww80} must be multiplied by the
same factor $\sqrt{2}/(3\sqrt{3})$. We have checked the matrix
elements (\ref{eq:tau1}), (\ref{eq:tau2}) also by direct computation
from the chiral Lagrangian (\ref{eq:L2}) (adding the appropriate
external charged vector field).

The amplitudes of $O(p^2)$ define the low-energy limit that all
amplitudes must satisfy in order to be consistent
with QCD. By themselves, they cannot be expected to provide a realistic
approximation to the physical amplitudes except in the immediate
threshold region. Naive extrapolation to the resonance region yields 
cross sections that are much smaller than the available experimental 
cross sections \cite{cmdl,cmdh}.

\section{Next-to-leading order}
\label{sec:p4} 
\renewcommand{\theequation}{\arabic{section}.\arabic{equation}}
\setcounter{equation}{0}
At $O(p^4)$, the amplitudes consist of the usual two parts: the first
one from one-loop diagrams with vertices of the lowest-order
Lagrangian (\ref{eq:L2}) and a second one from tree-level diagrams
with exactly one vertex of the chiral Lagrangian of $O(p^4)$:
\begin{equation}
A^\mu_{(4)} = A^\mu_{(4){\rm loop}} + A^\mu_{(4){\rm tree}}~.
\end{equation} 

\begin{figure}
\centerline{\epsfig{file=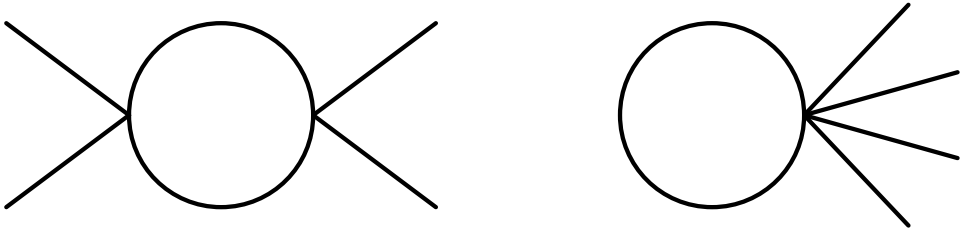,height=3cm}}
\caption{One-loop diagrams for $\gamma^* \to 4 \pi$. The virtual
photon is to be appended on all possible lines and vertices. Wave
function renormalization diagrams are not shown.}
\label{fig:loop}
\end{figure}
The loop amplitudes are calculated from diagrams of the general
form displayed in Fig.~\ref{fig:loop} where a virtual photon must be
appended wherever possible. For the loop
amplitude we have used a compact representation of the one-loop
generating functional for chiral $SU(2)$ with at most three 
propagators \cite{ru02} (of $O(\phi^6)$ in the notation of 
Refs.~\cite{gl84,gl85a}). We have checked the result in the limit of a
real photon ($q^2=0$) by comparing with the general formulas for
radiative four-meson amplitudes \cite{dein96}. Even the reduced
amplitude $A^\mu_{(4){\rm loop}}$ is quite lengthy
and we refer to Ref.~\cite{ru02} for the explicit form. Our excuse for
not reproducing it here is that that the one-loop amplitude will play a
relatively minor role for the cross sections in the experimentally
accessible region that we consider in this paper (0.65 $\le
E$(GeV) $\le$ 1.05).

The relevant part of the chiral $SU(2)$ Lagrangian of $O(p^4)$ is
given by \cite{gl84} (in the notation of, e.g., Ref.~\cite{egpr89})
\begin{eqnarray} 
{\cal L}_4 &=& \dfrac{l_1}{4}\langle u^\mu u_\mu \rangle^2 +
\dfrac{l_2}{4}\langle u_\mu u_\nu\rangle \langle u^\mu u^\nu \rangle + 
\dfrac{l_3}{16} \langle \chi_+\rangle^2 \nn
& & + \dfrac{i l_4}{4}\langle u_\mu\chi^\mu _- \rangle +
\dfrac{i l_6}{4}\langle f_+^{\mu\nu}[u_\mu,u_\nu]\rangle ~.
\label{eq:L4}
\end{eqnarray} 
The low-energy constants (LECs) $l_1,l_2$ appear in the (radiative) 
$\pi\pi$ amplitudes, $l_3, l_4$ enter through renormalization of the 
pion mass and decay constant and $l_6$ governs the
pion charge radius term in the expansion of the pion form factor. The
diagrams are the same as in Fig.~\ref{fig:tree} except that exactly
one vertex from (\ref{eq:L4}) must be inserted, with at most one other
vertex from the lowest-order Lagrangian (\ref{eq:L2}). The result
expressed in terms of the reduced amplitude $A^\mu_{(4){\rm tree}}$
is as follows:
\begin{eqnarray} 
F_\pi^4 A^\mu_{(4){\rm tree}}(p_1,p_2,p_3,p_4)& = & 
2 \wt l_2 (\nu-t_3) p_1^\mu \label{eq:tree4} \\
&& \hspace*{-4cm} + \left\{2 \wt l_1 (s^2-4 s M_\pi^2+ 4 M_\pi^4)+
\displaystyle\frac{\wt l_2}{2}(s^2 - 2 t_1 t_2 + 2 t_1^2 - 8 t_1 \nu
+ 4 \nu^2 - (q^2 - 2 t_3)^2) \right. \nn 
&& \left. + 2 \wt l_3 M_\pi^4 + 2 \wt l_4 (s M_\pi^2 - M_\pi^4) + 
\wt l_6 q^2 (M_\pi^2 - s) \right\} \displaystyle\frac{p_3^\mu}
{2 t_3 - q^2}~. \no
\end{eqnarray} 
We have included in $A^\mu_{(4){\rm tree}}$ the chiral logs from the
loop diagrams. The quantities $\wt l_i$, the amplitude 
(\ref{eq:tree4}) and the loop amplitude $A^\mu_{(4){\rm loop}}$
are then separately scale independent:
\begin{eqnarray} 
\wt l_1 = l_1^r(\mu) -
\displaystyle\frac{1}{96\pi^2}\ln{\displaystyle\frac{M_\pi^2}{\mu^2}},& 
~~~\wt l_2 = l_2^r(\mu) -
\displaystyle\frac{1}{48\pi^2}\ln{\displaystyle\frac{M_\pi^2}{\mu^2}},
& ~~~\wt l_3 = l_1^r(\mu) +
\displaystyle\frac{1}{64\pi^2}\ln{\displaystyle\frac{M_\pi^2}{\mu^2}},
\nn
\wt l_4 = l_4^r(\mu) -
\displaystyle\frac{1}{16\pi^2}\ln{\displaystyle\frac{M_\pi^2}{\mu^2}},&
~~~\wt l_6 = l_6^r(\mu) +
\displaystyle\frac{1}{96\pi^2}\ln{\displaystyle\frac{M_\pi^2}{\mu^2}} 
~. & \label{eq:lir}
\end{eqnarray}   

For the numerical analysis, we use the following values for the LECs
that correspond to the one-loop analysis in Ref.~\cite{cgl01}:
\begin{eqnarray} 
\hspace*{1cm} \wt l_1 = -2.0 \times 10^{-3} ~, & \hspace*{1cm} \wt l_2 = 1.1 
\times 10^{-2}~, & \hspace*{1cm} 
\wt l_3 = -4.6 \times 10^{-3}~, \nn
\hspace*{1cm} \wt l_4 = 2.8 \times 10^{-2}~, & \hspace*{1cm}
\wt l_6 = -1.7 \times 10^{-2}~.
\end{eqnarray} 

The $O(p^4)$ cross sections constructed from the amplitude
\begin{equation}
A^\mu = A^\mu_{(2)}+A^\mu_{(4){\rm loop}} + A^\mu_{(4){\rm tree}}
\label{eq:p4tot}
\end{equation} 
are shown as dotted curves in Figs.~\ref{fig:xsc},\ref{fig:xsn} for 
the energy range 
0.65 $\le E$(GeV) $\le$ 1.05. Comparison with the available data for the 
$2\pi^+ 2\pi^-$ channel \cite{cmdl} indicates that the theoretical
cross sections are still too small. The reason is clear: the
amplitudes of $O(p^4)$ contain only the low-energy remainders of resonance
exchange. We have to include meson resonance exchange explicitly if we
want to extrapolate the chiral amplitudes to the 1 GeV region.

\section{Resonance amplitudes generated at $\mbf{O(p^4)}$}
\label{sec:reso} 
\renewcommand{\theequation}{\arabic{section}.\arabic{equation}}
\setcounter{equation}{0}
The renormalized LECs $l_i^r(\mu)$ as well as their $SU(3)$
counterparts are known to be dominated by meson resonance exchange 
\cite{egpr89} at typical scales $\mu \sim M_\rho$. The tree-level
amplitude (\ref{eq:tree4}) of $O(p^4)$ therefore specifies 
how to extend the amplitude of $O(p^4)$ into the resonance region. 
In the $SU(3)$ notation, the relevant part of the resonance Lagrangian
is given by \cite{egpr89}:
\begin{eqnarray} 
\cL[V(1^{--}),A(1^{++}),S(0^{++})] &=& \cL_{\rm kin}[V,A,S,S_1]
\label{eq:resop4} \\
&& + \dfrac{F_V}{2 \sqrt{2}}\langle V_{\mu\nu} f_+^{\mu\nu}\rangle + 
    \dfrac{iG_V}{\sqrt{2}} \langle V_{\mu\nu} u^\mu u^\nu\rangle 
 + \dfrac{F_A}{2 \sqrt{2}} \langle A_{\mu\nu} f_-^{\mu\nu} \rangle 
\nn
&& + c_d  \langle S u_\mu u^\mu\rangle + 
                        c_m  \langle S \chi_+ \rangle +
   \wt c_d S_1  \langle u_\mu u^\mu \rangle +
    \wt c_m S_1  \langle \chi_+\rangle \no ~.
\end{eqnarray} 
The octets of vector and axial-vector mesons $V(1^{--}),A(1^{++})$ are
described by antisymmetric tensor fields $V_{\mu\nu},A_{\mu\nu}$ (see 
App.~\ref{app:VA}). $S,S_1$ are the 
scalar octet and singlet fields, respectively. Resonance exchange
dominance of the LECs at a scale $\mu=M_\rho$ amounts to the following
relations:
\begin{eqnarray} 
l_1^r(M_\rho) &=& - \displaystyle\frac{G_V^2}{M_\rho^2} +
2 \left( \displaystyle\frac{\wt c_d^2}{M_\sigma^2} +
\displaystyle\frac{c_d^2}{6 M_{f_0}^2}\right) \nn
l_2^r(M_\rho) &=& \displaystyle\frac{G_V^2}{M_\rho^2} \nn
l_3^r(M_\rho) &=& 8 \left( \displaystyle\frac{\wt c_m^2 -
\wt c_d \wt c_m}{M_\sigma^2} + \displaystyle\frac{c_m^2 -
c_d c_m}{6 M_{f_0}^2} \right) \nn
l_4^r(M_\rho) &=& 8 \left( \displaystyle\frac{\wt c_d \wt c_m}
{M_\sigma^2} + \displaystyle\frac{c_d c_m}{6 M_{f_0}^2} 
\right) \nn
l_6^r(M_\rho) &=& - \displaystyle\frac{F_V G_V}{M_\rho^2} ~.
\label{eq:lires}
\end{eqnarray} 
We have omitted the small contributions from kaon and eta loops
\cite{egpr89}. The axial coupling $F_A$ does not enter at this order
but will be needed in the following section. At the $SU(2)$ level,
there is of course no distinction between the $SU(2)$ singlet in $S$
and the $SU(3)$ singlet $S_1$. We associate the 
singlet field in $S$ with the $f_0$ and the $SU(3)$ singlet with the 
putative $\sigma$ meson. The overall contribution from scalar exchange
turns out to be very small so that the issue of scalar mixing
with or without glueballs \cite{ochs} has  no impact 
on our amplitudes in practice.

We use $M_\rho=0.775$ GeV and the following values for the vector 
couplings $F_V,G_V$:
\begin{eqnarray} 
F_V = 0.14 {\rm ~GeV}~, & \hspace*{2cm} G_V = 0.066 {\rm ~GeV} ~.
\label{eq:FVGV}
\end{eqnarray}
$G_V$ is obtained from the width $\Gamma (\rho \to \pi\pi)$= 0.15 GeV.
The chosen value for $F_V$ is the mean value of two possible
determinations from $\Gamma (\rho^0 \to e^+ e^-)$ and from the pion 
charge radius, respectively \cite{egpr89}. These values compare well
with the theoretically favoured values \cite{eglpr89}
\begin{eqnarray} 
F_V = \sqrt{2} F_\pi = 0.13 {\rm ~GeV}~, & \hspace*{1cm}  
G_V = F_\pi/\sqrt{2} =  0.065 {\rm ~GeV} ~.
\end{eqnarray}
In the scalar sector, we use \cite{egpr89}
\begin{eqnarray} 
c_d = 0.032 {\rm ~GeV}~, & \hspace*{1cm} c_m = 0.042 {\rm ~GeV}~, 
& \hspace*{1cm} \wt c_i = c_i/\sqrt{3} \hspace*{.6cm} (i= d,m)~,
\end{eqnarray}
with the latter nonet relation holding in the large-$N_c$ limit.

\begin{figure}[htb]
\centerline{\epsfig{file=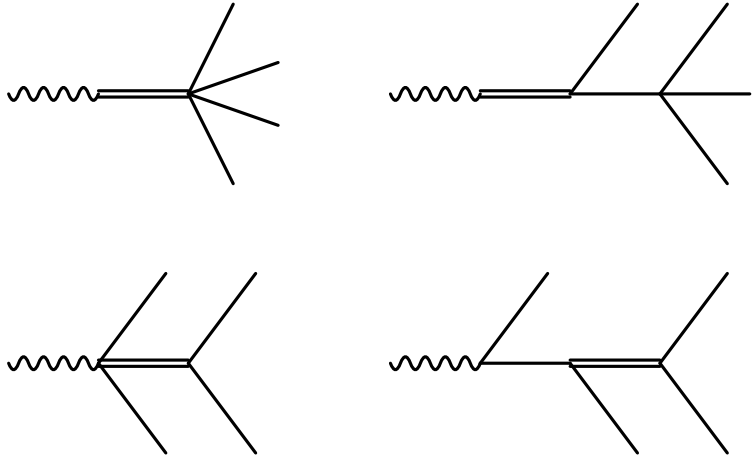,height=6cm}}
\caption{Resonance exchange diagrams contributing to the amplitudes
of $O(p^4)$. The double lines denote $\rho$, $\sigma$ and $f_0$ 
mesons.}
\label{fig:reso}
\end{figure}
In the tree-level amplitude (\ref{eq:tree4}) of $O(p^4)$
the renormalized LECs $l_i^r(M_\rho)$ are now set to zero and only the
chiral logs of Eq.~(\ref{eq:lir}) are kept. Instead, the explicit 
resonance exchange diagrams in Fig.~\ref{fig:reso} are calculated
giving rise to amplitudes $A_\rho^\mu$ and $A_S^\mu$, thereby
matching the $O(p^4)$ amplitude on to the resonance region:
\begin{eqnarray} 
F_\pi^4 A^\mu_\rho(p_1,p_2,p_3,p_4)& = & \nn
&& \hspace*{-2cm} 4 G_V^2 \left\{\displaystyle\frac{p_3^\mu p_1\cdot
p_2 - p_1^\mu p_2\cdot p_3}{D_\rho[(p_1+p_3)^2]} + 
\displaystyle\frac{2 p_3^\mu [p_1\cdot p_4 (p_2 \cdot p_3 - t_2)
- p_1\cdot p_2 (p_3 \cdot p_4 - t_4)]}{(2 t_3 -
q^2)D_\rho[(p_2+p_4)^2]} \right\} \nn
&& \hspace*{-2cm} + F_V G_V (p_3^\mu t_1 - p_1^\mu t_3)
\left\{\displaystyle\frac{1}{D_\rho(q^2)} - \displaystyle\frac{1}
{D_\rho[(p_1+p_3)^2]}\right\} \label{eq:rhop4} \\
&& \hspace*{-2cm} + \displaystyle\frac{F_V G_V q^2(s - M_\pi^2)} 
{D_\rho(q^2)(2 t_3 - q^2)} p_3^\mu \no
\end{eqnarray}
\begin{equation}  
F_\pi^4 A^\mu_S(p_1,p_2,p_3,p_4) = 
\displaystyle\sum_{S=f_0,\sigma} \displaystyle\frac
{2[(s - 2 M_\pi^2)c_d^S + 2 M_\pi^2 c_m^S]^2}
{3(2 t_3 - q^2) D_S(s)}p_3^\mu ~,
\end{equation}  
with scalar couplings
\begin{eqnarray} 
c_i^{f_0}=c_i ~, & \hspace*{1cm} c_i^{\sigma}= 
\sqrt{6} \,\wt c_i & \hspace*{1cm} (i=d,m)
\end{eqnarray} 
and propagators with energy dependent widths \cite{gdpp00}
\begin{eqnarray}  
D_P(t) &=& M_P^2 - t - i M_P \Gamma_P(t) \\
\Gamma_\rho(t) &=& \displaystyle\frac{M_\rho t}{96 \pi F_\pi^2}
(1 - 4 M_\pi^2/t)^{3/2}\theta(t-4 M_\pi^2) \label{eq:Grho} \\
\Gamma_S(t) &=& \Gamma_S \displaystyle\frac{t (1 - 4 M_\pi^2/t)^{1/2}}
{M_S^2 (1 - 4 M_\pi^2/M_S^2)^{1/2}}\Theta(t - 4 M_\pi^2)~.
\end{eqnarray}  
In the numerical analysis, we take 
\begin{eqnarray} 
M_{f_0}= 0.98 {\rm ~GeV}~, & \hspace*{1cm}  \Gamma_{f_0}=0.05 
{\rm ~GeV} ~, \nn
M_{\sigma}= 0.6 {\rm ~GeV}~, & \hspace*{1cm}  \Gamma_{\sigma}=0.6 
{\rm ~GeV} ~.
\end{eqnarray} 
Finally, we recall that $F_V G_V$ and $c_d c_m$ are both positive
\cite{egpr89}.
 
At this point, the amplitude $A^\mu(p_1,p_2,p_3,p_4)$ takes the
following form:
\begin{equation}
A^\mu = A^\mu_{(2)} + A^\mu_{(4){\rm loop}} + \wh A^\mu_{(4){\rm tree}}
+ A^\mu_\rho + A^\mu_S 
\label{eq:p4reso}
\end{equation} 
where the amplitude $\wh A^\mu_{(4){\rm tree}}$ contains only the
chiral logs in (\ref{eq:lir}) for $\mu=M_\rho$. The renormalized 
LECs $l_i^r(M_\rho)$ have been traded for the explicit resonance
exchange amplitudes $A^\mu_\rho, A^\mu_S$. The resulting cross 
sections are more realistic than the
strictly $O(p^4)$ cross sections from the amplitude (\ref{eq:p4tot}),
but they are still too
small in comparison with the available data around 1 GeV 
\cite{cmdl,cmdh}.

\section{Beyond $\mbf{O(p^4)}$}
\label{sec:beyond} 
\renewcommand{\theequation}{\arabic{section}.\arabic{equation}}
\setcounter{equation}{0}

There must be additional ingredients in the amplitudes that make 
important contributions to the cross sections already for energies 
below 1 GeV. To the extent that the LECs $l_i^r(M_\rho)$ are known to
be dominated by $\rho$ (and scalar) exchange as given in 
Eqs.~(\ref{eq:lires}), those additional amplitudes must
vanish\footnote{Small additional contributions to the LECs of 
$O(p^4)$ are possible and even expected, e.g. from $\rho^\prime$
exchange.} to $O(p^4)$. We will try to locate the dominant
contributions that appear first at $O(p^6)$ but in contrast to the
previous section we cannot claim completeness here. At this order,
also diagrams with more than one meson resonance 
contribute. Even for single resonance exchange, a complete analysis 
of $O(p^6)$ is not available at present.

However, we may turn to existing phenomenological treatments 
\cite{dhjf96,ck01,novosibirsk} for guidance. In addition to the
obvious $\rho$ (and the less important scalar) exchange, the data
\cite{cmdl,cmdh} clearly indicate the presence of $\omega$ and $a_1$ 
exchange. 

\subsection{Double $\mbf{\rho}$ exchange}
\label{sec:2rho} 
\begin{figure}
\centerline{\epsfig{file=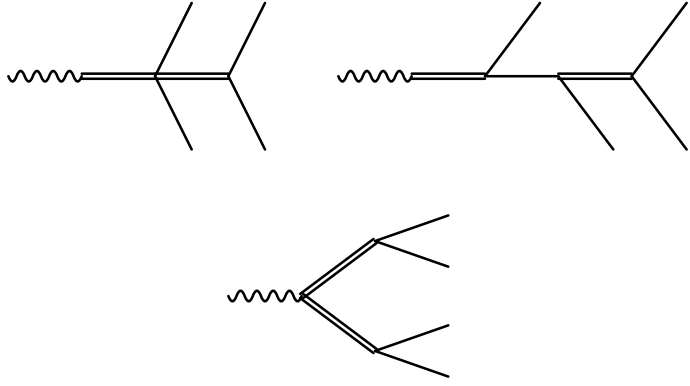,height=5cm}}
\caption{Double $\rho$ exchange diagrams generated by the Lagrangian
(\protect\ref{eq:resop4}).}
\label{fig:2rho}
\end{figure}
The resonance Lagrangian (\ref{eq:resop4}) also generates amplitudes 
starting at $O(p^6)$ with two $\rho$ mesons exchanged. The
corresponding diagrams are displayed in Fig.~\ref{fig:2rho}. 
The diagram where the virtual photon couples to $\rho^+ \rho^-$ is 
actually required by gauge invariance because the (charged) vector
mesons are dynamical fields here. The diagrams of Fig.~\ref{fig:reso}
produce a gauge invariant amplitude in the strict $O(p^4)$ limit only
where the resonance propagators shrink to points. Although of different
chiral order, the amplitudes of Figs.~\ref{fig:reso} and
\ref{fig:2rho} must be added for a meaningful amplitude in the
resonance region.

All couplings needed for the diagrams of Fig.~\ref{fig:2rho} have
already been defined. The (reduced) double $\rho$ exchange amplitude
has the explicit form
\begin{eqnarray} 
F_\pi^4 A^\mu_{\rho\rho}(p_1,p_2,p_3,p_4)& = & \nn
&& \hspace*{-3cm} \displaystyle\frac{4 G_V^2}{D_\rho[(p_1+p_3)^2]
D_\rho[(p_2+p_4)^2]}\left\{p_2^\mu(M_\pi^2+p_1\cdot p_3)(p_1 - p_3)
\cdot p_4 \right.  \nn
&&  \hspace*{2.1cm} \left. + p_3^\mu(M_\pi^2+p_2\cdot p_4)p_1\cdot 
(p_2 - p_4)\right\} + \label{eq:2rho} \\ 
&& \hspace*{-3cm} \displaystyle\frac{F_V G_V}{D_\rho(q^2)
D_\rho[(p_2+p_4)^2]}\left\{p_2^\mu q^2(p_1 - p_3)\cdot p_4+
p_4^\mu q^2 p_2\cdot (p_3 - p_1)\right. \nn
&& \hspace*{-2cm} + \left.[(p_1 - p_3)^\mu(t_4 - t_2) -
(p_2 - p_4)^\mu(t_3 - t_1)](M_\pi^2+p_2\cdot p_4)\right\} + \nn
&& \hspace*{-3cm} \displaystyle\frac{8 F_V G_V^3 q^2}{F_\pi^2 
D_\rho(q^2)D_\rho[(p_2+p_4)^2](2 t_3 - q^2)} p_3^\mu \left
\{p_1\cdot p_4 (p_2\cdot p_3 - t_2) - p_1\cdot p_2 
(p_3\cdot p_4 - t_4)\right\} ~.\no
\end{eqnarray} 

Putting together the lowest-order amplitude $A^\mu_{(2)}$ in
(\ref{eq:Ap2}) and the single and double $\rho$ exchange amplitudes
$A^\mu_\rho$, $A^\mu_{\rho\rho}$ in (\ref{eq:rhop4}), 
(\ref{eq:2rho}), one observes that some terms can be combined as the
leading terms in the low-energy expansion of the pion form factor
in view of the relation $F_V G_V \simeq F_\pi^2$ \cite{eglpr89}:
\begin{equation} 
F_\pi(q^2) = 1 + \displaystyle\frac{F_V G_V q^2}{F_\pi^2 D_\rho(q^2)}
+ \displaystyle\frac{i}{M_\rho} \Gamma_\rho(q^2) + \dots \simeq
\displaystyle\frac{M_\rho^2}{D_\rho(q^2)} ~.
\label{eq:FVq2}
\end{equation}
We have included the leading-order absorptive part 
$i \Gamma_\rho(q^2)/M_\rho$ that is contained in the one-loop
amplitude $A^\mu_{(4){\rm loop}}$ in one case and is of higher order
in the other case. Replacing the
expansion terms on the left-hand side of (\ref{eq:FVq2}) by the
usual $\rho$ dominance form (the right-hand side of
Eq.~(\ref{eq:FVq2})) is equivalent to the order considered. Of course,
the partial resummation yields a phenomenologically much more
realistic amplitude. A similar resummation applies to the scalar
exchange amplitude. We therefore express the combined $\rho$ and
scalar exchange amplitude in the following form 
\begin{equation}
 \left(A^\mu_{(2)} + A^\mu_S\right) M_\rho^2/D_\rho(q^2) 
+ \wh A^\mu_\rho + \wh A^\mu_{\rho\rho} ~. \label{eq:Arho}
\end{equation} 
The modified single
and double $\rho$ exchange amplitudes are now given by
\begin{eqnarray} 
F_\pi^4 \wh A^\mu_\rho(p_1,p_2,p_3,p_4)& = & \nn
&& \hspace*{-3cm} 4 G_V^2 \left\{\displaystyle\frac{p_3^\mu p_1\cdot
p_2 - p_1^\mu p_2\cdot p_3}{D_\rho[(p_1+p_3)^2]} + 
\displaystyle\frac{2 p_3^\mu [p_1\cdot p_4 (p_2 \cdot p_3 - t_2)
- p_1\cdot p_2 (p_3 \cdot p_4 - t_4)]M_\rho^2}{(2 t_3 -
q^2)D_\rho[(p_2+p_4)^2] D_\rho(q^2)} \right\} \nn
&& \hspace*{-3cm} + F_V G_V (p_3^\mu t_1 - p_1^\mu t_3)
\left\{\displaystyle\frac{1}{D_\rho(q^2)} - \displaystyle\frac{1}
{D_\rho[(p_1+p_3)^2]}\right\} \label{eq:rhop4mod} 
\end{eqnarray}
\begin{eqnarray} 
F_\pi^4 \wh A^\mu_{\rho\rho}(p_1,p_2,p_3,p_4)& = & \nn
&& \hspace*{-3cm} \displaystyle\frac{4 G_V^2}{D_\rho[(p_1+p_3)^2]
D_\rho[(p_2+p_4)^2]}\left\{p_2^\mu(M_\pi^2+p_1\cdot p_3)(p_1 - p_3)
\cdot p_4 \right.  \nn
&&  \hspace*{2.1cm} \left. + p_3^\mu(M_\pi^2+p_2\cdot p_4)p_1\cdot 
(p_2 - p_4)\right\} +  \nn 
&& \hspace*{-3cm} \displaystyle\frac{F_V G_V}{D_\rho(q^2)
D_\rho[(p_2+p_4)^2]}\left\{p_2^\mu q^2(p_1 - p_3)\cdot p_4+
p_4^\mu q^2 p_2\cdot (p_3 - p_1)\right. \nn
&& \hspace*{-2cm} + \left.[(p_1 - p_3)^\mu(t_4 - t_2) -
(p_2 - p_4)^\mu(t_3 - t_1)](M_\pi^2+p_2\cdot p_4)\right\} ~.
\label{eq:2rhomod}
\end{eqnarray} 
The leading-order absorptive part of the pion form factor must 
be subtracted from the one-loop amplitude yielding a modified 
loop amplitude $\wh A^\mu_{(4){\rm loop}}$.

\subsection{$\mbf{\omega}$ exchange: vector meson dominance}
\label{sec:omega} 
Vector meson dominance (VMD) for $\omega$ decays postulates the
dominant role of an $\omega \rho \pi$ coupling \cite{gmsw62,sak69}.
We write the corresponding Lagrangian (unique to lowest order in 
derivatives) as
\begin{equation} 
\cL(\omega\rho\pi)=g_{\omega\rho\pi}\ve_{\mu\nu\rho\sigma}
\omega^\mu \partial^\nu \vec{\pi} \cdot \vec{\rho}^{\rho\sigma}~.
\end{equation}
In this case, it is more convenient to describe the $\omega$ in 
terms of a vector field $\omega^\mu$ (see App.~\ref{app:VA}). 

The decay $\omega \to \rho^0 \pi^0 \to \pi^0 \gamma$ proceeds with
a rate
\begin{equation} 
\Gamma(\omega \to \pi^0 \gamma) = \displaystyle\frac
{\alpha g_{\omega\rho\pi}^2 F_V^2}{6 M_\omega^3 M_\rho^4}
\left(M_\omega^2 - M_\pi^2\right)^3 ~. 
\end{equation} 
The measured partial width \cite{pdg00} corresponds to 
$|g_{\omega\rho\pi}|=5.0$. On the other hand, the dominant decay chain
$\omega \to \rho \pi \to 3\pi$ leads to $|g_{\omega\rho\pi}|=5.7$.
A small direct $\omega \to 3\pi$ amplitude is allowed but VMD clearly 
accounts for the dominant features of both decays.

\begin{figure}[htb]
\centerline{\epsfig{file=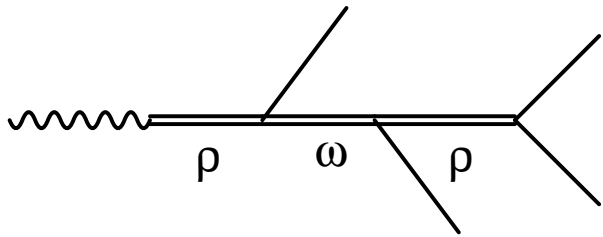,height=3cm}}
\caption{VMD diagram for $\omega$ exchange.}
\label{fig:omega}
\end{figure}

For $e^+ e^- \to 4 \pi$, the relevant $\omega$ exchange diagram is shown in
Fig.~\ref{fig:omega}. 
It gives rise to a (reduced) amplitude
\begin{eqnarray}  
A^\mu_{\omega}(p_1,p_2,p_3,p_4) &=& 
\displaystyle\frac{8 F_V G_V g_{\omega\rho\pi}^2}
{F_\pi^2 D_\rho(q^2) D_{\omega}[(q-p_1)^2]}
\left\{- p_2^\mu t_4 p_1\cdot p_3 +
p_3^\mu(t_4 p_1\cdot p_2 - t_2 p_1\cdot p_4)\right \} \nn
&&\left\{D_\rho^{-1}[(p_2+p_3)^2] + D_\rho^{-1}[(p_2+p_4)^2] +
D_\rho^{-1}[(p_3+p_4)^2]\right\} ~.
\label{eq:omex}
\end{eqnarray} 
In view of the small value $\Gamma_\omega =$ 8.44 MeV \cite{pdg00} we
employ an energy independent width in the propagator
function $D_{\omega}[(q-p_1)^2]$. The amplitude (\ref{eq:omex})
completely dominates the cross section
for $e^+ e^- \to 2\pi^0\pi^+\pi^-$ around 1 GeV in accordance with
experimental findings \cite{cmdh,novosibirsk}. In order to appreciate
the size of this amplitude of $O(p^6)$, we compare it to a typical 
$\rho$ exchange amplitude of $O(p^4)$ as given in
(\ref{eq:rhop4}). By naive chiral counting, the dimensionless
quantity $c_\omega$ defined by
\begin{equation} 
\displaystyle\frac{c_\omega}{(4\pi F_\pi)^2} = 
\displaystyle\frac{24 g_{\omega\rho\pi}^2 F_\pi^2}{M_\omega^2 
M_\rho^2}
\end{equation}   
would be expected to be of $O(1)$. With $|g_{\omega\rho\pi}|=5.7$
one finds instead 
$|c_\omega|=24$, quite a drastic deviation from naive chiral counting.
The sign of the $\omega$ exchange amplitude (\ref{eq:omex}) is fixed
by the positive sign of $F_V G_V$ \cite{egpr89,eglpr89}.
Of course, the corresponding amplitude due to $\phi$ exchange is
completely negligible.

\subsection{$\mbf{a_1}$ exchange}
\label{sec:a1} 
Although $\omega$ exchange dominates the cross section for the 
$2 \pi^0 \pi^+ \pi^-$ final state it does not contribute to the other
channel $2 \pi^+ 2 \pi^-$. Here $a_1$ exchange will play an important
role. We follow the usual VMD assumption that the dominant decay mode 
$a_1 \to 3 \pi$ proceeds via an intermediate $\rho$.

\begin{figure}[htb]
\centerline{\epsfig{file=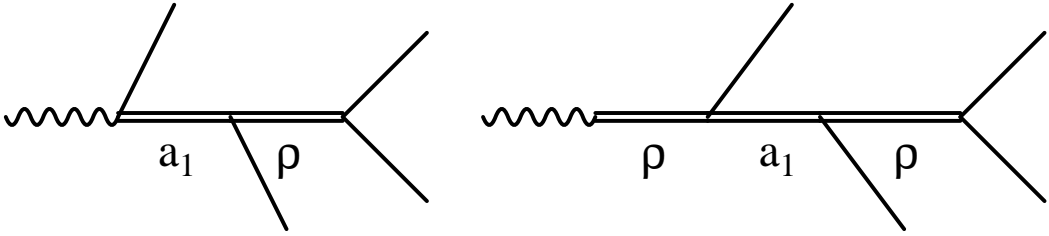,height=3cm}}
\caption{$a_1$ exchange diagrams.}
\label{fig:a1}
\end{figure}

Contrary to $\omega \rho \pi$, there are several possible chiral 
couplings for the $a_1 \rho \pi$ vertex.
The ideal place to analyse this vertex is the process 
$\tau \to \nu_\tau 3 \pi$ and such an analysis is under way \cite{pp02}.
In the tensor field formalism, there are altogether five a priori 
independent $a_1 \rho \pi$ couplings of lowest possible chiral order
\cite{pp02}. Two of them give the same amplitudes in our case and
another one is proportional to $M_\pi^2$ and therefore
vanishes in the chiral limit \cite{pp02}. We will restrict ourselves
here to the remaining terms that boil down to the following Lagrangian
for the charged $a_1$ fields (the neutral $a_1$ cannot be exchanged in
the diagrams of Fig.~\ref{fig:a1}):
\begin{eqnarray} 
\cL (a_1^+ \to \rho \pi) &=&  \displaystyle\frac{i c_2}{M_{a_1}} 
a_1^{+\mu\nu}\left(\rho^0_{\nu\lambda}
\partial_\mu\partial^\lambda \pi^- - \rho^-_{\nu\lambda}
\partial_\mu\partial^\lambda \pi^0\right) \nn
&+&  \displaystyle\frac{i c_3}{M_{a_1}} a_1^{+\mu\nu}
\left(\partial^\lambda \rho^0_{\nu\lambda}
\partial_\mu \pi^- - \partial^\lambda \rho^-_{\nu\lambda}
\partial_\mu \pi^0\right) \nn
&+& \displaystyle\frac{i c_4}{M_{a_1}} a_1^{+\mu\nu}
\left(\partial_\mu \rho^0_{\lambda\nu}
\partial^\lambda \pi^- - \partial_\mu \rho^-_{\lambda\nu}
\partial^\lambda \pi^0\right) + {\rm~h.c.}~,
\label{eq:a1rhopi}
\end{eqnarray} 
with dimensionless couplings $c_2,c_3,c_4$.

The analysis of $\tau \to \nu_\tau 3 \pi$ 
that should determine or at least relate the constants
$c_i$ is not yet available \cite{pp02}.
In order to proceed, we make the simplifying assumption that
the couplings are all equal:
\begin{eqnarray} 
&c_2 = c_3 = c_4 ~.&
\label{eq:ci}
\end{eqnarray}
From the decay width $\Gamma(a_1 \to \rho \pi \to 3 \pi)$, accounting
for the finite $\rho$ width, we find
\begin{equation} 
|c_2| = 319 ~[\Gamma(a_1 \to 3 \pi)/0.5 {\rm ~GeV}]^{1/2} ~.
\end{equation}
The surprisingly large value of $|c_2|$ is due to the fact that the
(off-shell) $a_1 \to \rho \pi \to 3 \pi$ vertex function vanishes in 
the chiral limit for the choice (\ref{eq:ci}), as long as the pions are
on-shell. Although this
property is certainly not required by chiral symmetry it has
interesting implications for the high-energy behaviour of the
$\tau \to \nu_\tau 3 \pi$ amplitude \cite{pp02}. In this case, the
neglected coupling proportional to $M_\pi^2$ should be added in a
complete analysis of the $a_1 \rho \pi$ vertex. We have also
investigated other choices for the couplings $c_i$: the resulting
cross sections are always smaller than for the choice
(\ref{eq:ci}).

As shown in Fig.~\ref{fig:a1}, there are two $a_1$ exchange diagrams
that must be taken into account here. The first one has a direct
$a_1 \pi \gamma$ coupling $F_A$ defined in the resonance
Lagrangian (\ref{eq:resop4}). We take the theoretically favoured value
\cite{eglpr89} $F_A = F_\pi$ for this coupling. In addition, also the
$a_1 \rho \pi$ couplings in (\ref{eq:a1rhopi}) contribute to the
radiative decay $a_1 \to \pi \gamma$. For the choice (\ref{eq:ci}) we
obtain an effective coupling 
\begin{equation} 
F_A^{\rm eff} = F_A - \displaystyle\frac{c_2 F_V F_\pi M_\pi^2}
{M_{a_1} M_\rho^2} \label{eq:FAeff} ~,
\end{equation}
with the two terms approximately equal in magnitude. Requiring
constructive interference (sgn$(c_2 F_A F_V F_\pi) < 0$), the
resulting partial width $\Gamma(a_1 \to \pi \gamma)$ is larger than the
PDG value of 640 keV (based on a single experiment)  by about a factor 
2.5 for $\Gamma(a_1 \to 3 \pi)=$ 0.5 GeV.

The (reduced) amplitude $A^\lambda_{a_1}(p_1,p_2,p_3,p_4)$ from the 
two diagrams in Fig.~\ref{fig:a1} is in an obvious
notation given by
\begin{eqnarray}
A^\lambda_{a_1}(p_1,p_2,p_3,p_4) &=& 
\displaystyle\frac{V_L^{\lambda\mu\nu} N_{\mu\nu\rho\sigma}
(q - p_4) V_R^{\rho\sigma}}{M_{a_1}^2 D_{a_1}[(q-p_4)^2]
D_{\rho}[(p_1+p_3)^2]} \label{eq:a1ex}\\
V_L^{\lambda\mu\nu} &=& \displaystyle\frac{F_V}{M_{a_1}
D_\rho(q^2)}\left[c_2 p_4^\mu(t_4 g^{\lambda\nu}-p_4^\lambda q^\nu)
- c_3 q^2 p_4^\mu g^{\lambda\nu} + c_4 t_4 q^\mu g^{\lambda\nu}
\right] \nn
&& - \displaystyle\frac{F_A}{F_\pi} q^\mu g^{\lambda\nu} \nn
N_{\mu\nu\rho\sigma}(k) &=&
g_{\mu \rho} g_{\nu\sigma}(M_{a_1}^2-k^2) 
 + g_{\mu\rho} k_\nu k_\sigma -
g_{\mu\sigma} k_\nu k_\rho - (\mu \leftrightarrow \nu) \nn
V_R^{\rho\sigma} &=& \displaystyle\frac{2 G_V}{F_\pi^2 M_{a_1}}
\left \{ c_2(p_1\cdot p_2 p_2^\rho p_3^\sigma - 
p_2\cdot p_3 p_2^\rho p_1^\sigma) 
+ c_3 (p_1\cdot p_3 + M_\pi^2)p_2^\rho (p_3^\sigma -  p_1^\sigma)
\right. \nn & & +\left. c_4 p_2\cdot (p_1 + p_3)
p_3^\rho p_1^\sigma \right\} ~. \no 
\end{eqnarray}
The relative signs in this amplitude are determined by the choice
(\ref{eq:ci}), the constructive interference in (\ref{eq:FAeff}) and
by $F_V G_V > 0$ \cite{egpr89,eglpr89}.

The energy dependence of the $a_1$ width $\Gamma_{a_1}(t)$ has also 
a considerable numerical impact. Awaiting the results of 
Ref.~\cite{pp02}, we assume for the present analysis the functional
form (all numbers are to be understood in appropriate units
of GeV) suggested by K\"uhn and Santamaria \cite{ks90}:
\begin{eqnarray}
\Gamma_{a_1}(t) &=& \Gamma_{a_1} g(t)/g(M_{a_1}^2) \nn 
g(t) &=& (1.623 \,t + 10.38 - 9.23/t + 0.65/t^2) 
\,\Theta[t - (M_\rho + M_\pi)^2] \\
&& \hspace*{-1.5cm} + ~4.1 (t - 9 M_\pi^2)^3 [1 - 3.3 (t - 9 M_\pi^2)
+ 5.8 (t - 9 M_\pi^2)^2 ]\,\Theta(t - 9 M_\pi^2)\Theta[(M_\rho +
M_\pi)^2 - t] \no ~.
\end{eqnarray} 

\begin{figure}[ht]
\setlength{\unitlength}{1cm}
\begin{picture}(16,8)
\put(0.5,0.03){\makebox(7.0,7.0)[lb]{\epsfig
{figure=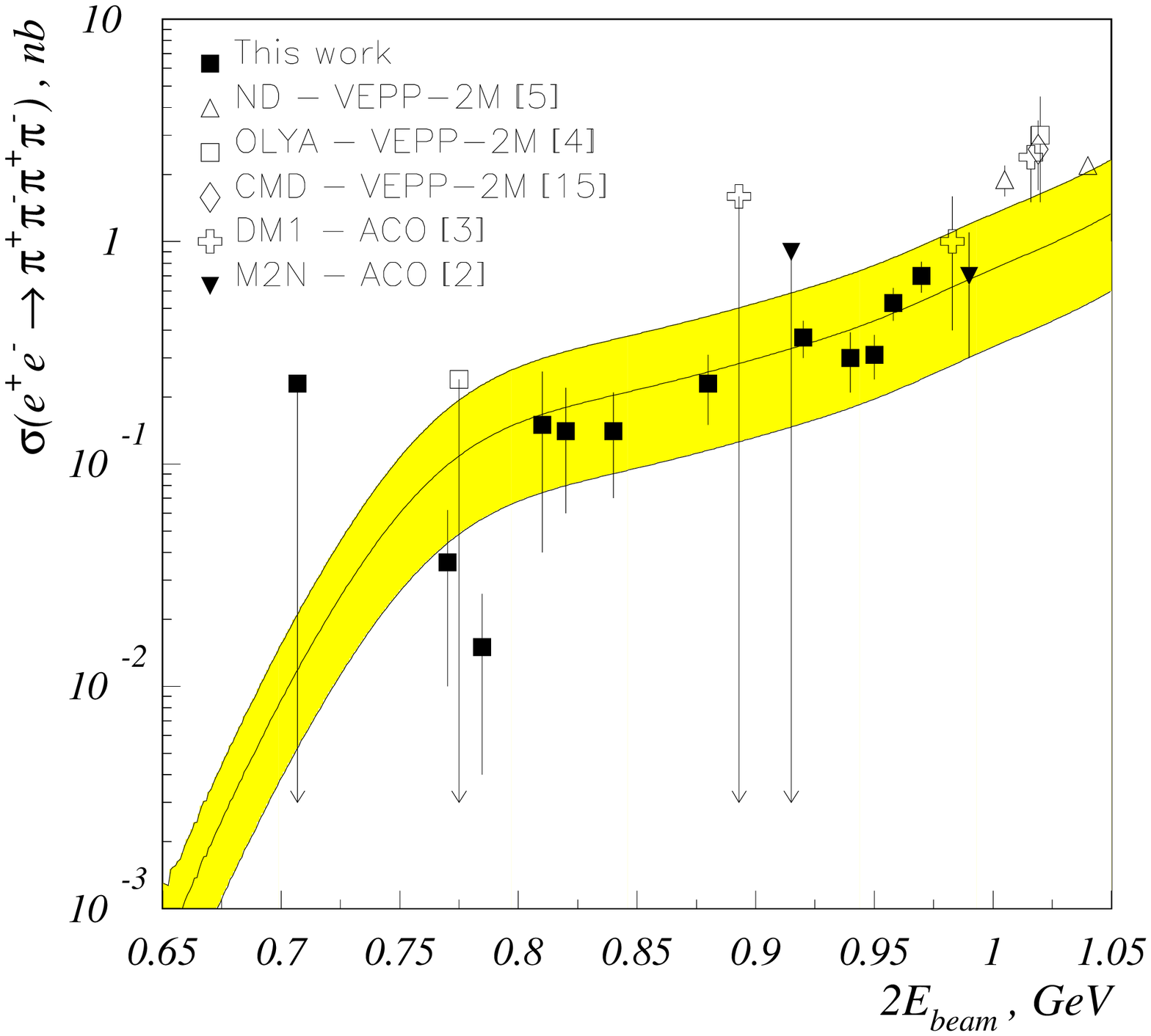,width=6.56cm,height=7.07cm}}}
\put(8.5,0.6){\makebox(7.0,7.0)[lb]
{\epsfig{figure=4pxsc4.eps,height=6cm}}}
\end{picture}
\caption{Comparison of data \protect\cite{cmdl} (left figure) and 
predictions (right figure, see text) for the cross section 
$\sigma(e^+ e^- \to 2\pi^+ 2\pi^-)$ for 0.65 $\le E$(GeV) $\le$ 1.05.} 
\label{fig:xsc}
\end{figure}

Putting all contributions together, we arrive at our final amplitude
\begin{equation}
A^\mu_{\rm final} = \left\{A^\mu_{(2)}   
+ A^\mu_S \right\} M_\rho^2/D_\rho(q^2) 
+ \wh A^\mu_{(4){\rm loop}} + \wh A^\mu_{(4){\rm tree}}
+ \wh A^\mu_\rho + \wh A^\mu_{\rho\rho}
+  A^\mu_\omega +  A^\mu_{a_1} ~. \label{eq:Afinal}
\end{equation} 
This amplitude has the correct low-energy behaviour to $O(p^4)$ and is
expected to contain the relevant ingredients for an
extrapolation to the 1 GeV region.

\section{Cross sections and decay rates}
\label{sec:rates} 
\renewcommand{\theequation}{\arabic{section}.\arabic{equation}}
\setcounter{equation}{0}
In Fig.~\ref{fig:xsc} we compare our results for the cross section
$\sigma(e^+ e^- \to 2 \pi^+ 2\pi^-)$ with available data taken from
Ref.~\cite{cmdl}. The dotted curve corresponds to the strictly
$O(p^4)$ amplitude (\ref{eq:p4tot}) and the full curve is the cross 
section for the final amplitude (\ref{eq:Afinal}).
Whereas the dotted curve is definitely too low 
the full curve agrees well with experimental 
data up to 1 GeV. The more pronounced rise of the full 
curve is mainly due to $a_1$ exchange. The $\rho$ exchange amplitude 
generated at $O(p^4)$ is also important, to a lesser extent also the
lowest-order amplitude with resummed pion form factor. Loops and
chiral logs are much less important. Finally, scalar 
exchange contributes very little to the cross section and 
double $\rho$ exchange does not contribute at all in this channel. 

At low energies where our amplitude should be most reliable the
predicted cross section is below the shaded band in Fig.~\ref{fig:xsc}. 
That band \cite{cmdl} describes an extrapolation from data at higher
energies \cite{cmdh} with a non-chiral resonance model.

\begin{figure}
\centerline{\epsfig{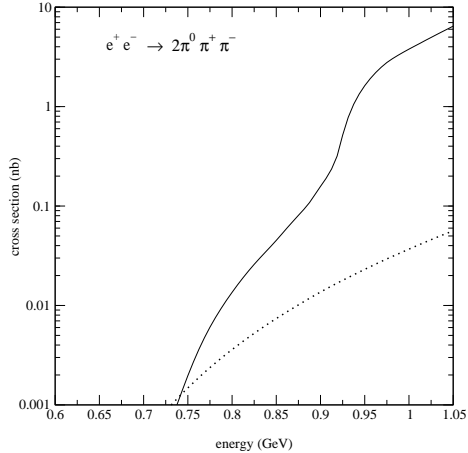}}
\caption{Theoretical predictions for the cross section $\sigma(e^+ e^- \to 
2\pi^0\pi^+\pi^-)$ for 0.65 $\le E$(GeV) $\le$ 1.05.}
\label{fig:xsn}
\end{figure}
The analogous theoretical results for $\sigma(e^+ e^- \to 2 \pi^0
\pi^+ \pi^-)$ are shown in Fig.~\ref{fig:xsn}. Although there are so
far no data in the region below 1 GeV the theoretical cross section
(full curve) connects well with the data starting at 1 GeV 
\cite{cmdh}. For this channel, the much steeper rise compared to the
$2 \pi^+ 2\pi^-$ mode is almost exclusively due to $\omega$ exchange.
$a_1$ exchange, even though less important here, interferes
constructively with $\omega$ exchange in the 1 GeV region. All 
other contributions are very small there.

Near the $\rho$ pole, the amplitudes are of course completely
dominated by the resonant parts containing the propagator function 
$D_\rho^{-1}(q^2)$. The cross sections at $E=M_\rho$ therefore
determine the branching ratios for
$\rho^0 \to 4 \pi$ according to the general formula
\begin{equation} 
BR(\rho^0 \to f)=\displaystyle\frac{M_\rho^2 ~\sigma
(e^+ e^- \to f)|_{E=M_\rho}}{12 \pi ~BR(\rho^0 \to e^+ e^-) } ~. 
\end{equation} 
For the channel $2 \pi^+ 2\pi^-$, the relevant contributions are
the lowest-order amplitude with pion form factor, single $\rho$
and $a_1$ exchange. There is a destructive interference between the 
modified lowest-order amplitude on the one hand and the two 
single-resonance exchange amplitudes at $E=M_\rho$. Except for $a_1$
exchange, this interference is dictated by the QCD structure
of $O(p^4)$. Although the $a_1$ amplitude depends on our assumption
(\ref{eq:ci}) for the $a_1\rho\pi$ vertex the relative sign to the 
other two amplitudes is also fixed. The situation is a little 
different in the $2 \pi^0 \pi^+ \pi^-$ channel because of the 
additional $\omega$ exchange. In this channel, the interference 
pattern is: lowest-order amplitude with pion form factor + $a_1$ 
exchange -- single $\rho$ exchange -- $\omega$ exchange.
Although the couplings involved
are relatively well known we assign a 40 $\%$ error to the calculated
branching ratios in view of the destructive interferences:
\begin{eqnarray} 
BR(\rho^0 \to 2 \pi^+ 2\pi^- ) &=& (6.7 \pm 2.7)\times 10^{-6} \nn
BR(\rho^0 \to 2 \pi^0 \pi^+ \pi^- ) &=& (5.0 \pm 2.0)\times 10^{-6}~.
\label{eq:rhowidth}
\end{eqnarray} 
The result for the $2 \pi^+ 2\pi^-$ mode agrees within errors with the 
experimental value \cite{pdg00} (extracted from the cross section in 
Fig.~\ref{fig:xsc} at $E=M_\rho$) although our mean value
is almost a factor three smaller. For the 
$2 \pi^0 \pi^+ \pi^-$ channel there is only an experimental upper
limit \cite{pdg00} that is compatible with (\ref{eq:rhowidth}). There
is a wide range of model predictions for the $4 \pi$ decay modes
of the $\rho$ as reviewed in Ref.~\cite{pb96}.

We have plotted the cross sections only for energies $\lets$ 1 GeV
because our amplitudes do not have the correct high-energy 
behaviour. This manifests itself already in the 1 to 2 GeV
region where our cross sections exceed the
experimental cross sections \cite{cmdh}.

Scaling all four-momenta in the same way and requiring that 
$\sigma(e^+ e^- \to 4 \pi)$ decreases at least as fast as (most likely
faster than) $1/E^2$ at 
large energies to satisfy the asymptotic QCD constraint, one finds
that the basic current matrix element $J^\mu(p_1,p_2,p_3,p_4)$ in
(\ref{eq:k1}) must vanish at large energies at least as $1/E$. This
criterion is not even met by the lowest-order matrix element
(\ref{eq:Jp2}) although it is satisfied by the modified lowest-order
amplitude in (\ref{eq:Arho}) due to the pion form factor. 

In addition to resummations of parts of the amplitude, additional
higher-mass states must be included in order to access the region up
to 2 GeV and to satisfy the high-energy constraints of QCD. The
Particle Data Group lists \cite{pdg00} many such resonances with the 
appropriate quantum numbers, e.g., $\rho(1450)$, $\rho(1700)$,
$f_0(1370)$, $f_0(1500)$, $f_0(1710)$, $\omega(1420)$, $\omega(1650)$ 
and states with higher spins. In the spirit of duality, all
those states are expected to conspire to produce the right 
asymptotic behaviour of the amplitudes at high energies.

\section{Conclusions}
\label{sec:conc} 
\renewcommand{\theequation}{\arabic{section}.\arabic{equation}}
\setcounter{equation}{0}
We have performed the first calculation of the processes 
$e^+ e^- \to  4\pi$ and $\tau \to \nu_\tau 4 \pi$ with the correct
structure to $O(p^4)$ in the low-energy expansion of the standard
model. In addition to the proper low-energy structure, CHPT
automatically produces amplitudes with all relevant symmetries of 
the standard model, 
in our case (spontaneously and softly broken) chiral symmetry, gauge
invariance, Bose symmetry and $C$ invariance. 

Although the chiral amplitude to $O(p^4)$ is only valid close to
threshold it contains information how to extrapolate to the
resonance region. This information on $\rho$ and scalar exchange is
however not sufficient to describe the $e^+ e^-$ cross sections up
to energies of around 1 GeV. We have therefore included as additional
contributions $\omega$, $a_1$ and double $\rho$ exchange that
first show up at $O(p^6)$. All necessary couplings were determined
from the decay widths of the various resonances involved.

The predicted cross sections for $E \lets$ 1 GeV and the branching
ratios $BR(\rho^0 \to 4 \pi)$ are in good agreement with 
available data. Our amplitudes do not have an acceptable high-energy
behaviour so that additional ingredients are needed (such as
higher-mass resonance exchange) to make predictions in the
phenomenologically interesting region up to 2 GeV. 

In the isospin limit, the $\tau$ decay amplitudes can be calculated
from the annihilation amplitude for the channel $2 \pi^0 \pi^+ \pi^-$
\cite{kuehnrev}. The comparison with $\tau$ decay data will be postponed
until the proper high-energy behaviour has been implemented.

\vfill
\section*{Acknowledgements}
\noindent 
We thank J. Portol\'{e}s for informing us of his unpublished work with
A. Pich on the $a_1 \rho \pi$ vertex and J.H. K\"uhn for helpful 
informations.

\medskip\medskip
\newcounter{zaehler}
\renewcommand{\thesection}{\Alph{zaehler}}
\renewcommand{\theequation}{\Alph{zaehler}.\arabic{equation}}
\setcounter{zaehler}{1}
\setcounter{equation}{0}

\section{Isospin relations}
\label{app:iso}
From the isospin relations \cite{kuehnrev} 
(\ref{eq:k1}),\dots,(\ref{eq:k4}) for the four possible final states 
it is evident that the amplitude for the $2 \pi^0 \pi^+ \pi^-$
channel is sufficient to obtain the remaining three amplitudes.
From the observation that $\omega$ exchange cannot contribute to
the $2 \pi^+ 2 \pi^-$, $3 \pi^0 \pi^-$ modes one finds immediately
that none of those latter modes is sufficient to calculate the
remaining ones.

The only nontrivial question\footnote{We thank Hans K\"uhn for raising
this question.} concerns the channel $2 \pi^- \pi^+ \pi^0$. At first
sight, one would expect that from the sum of two current matrix
elements in
(\ref{eq:k3}) one cannot determine $J^\mu(p_1,p_2,p_3,p_4)$
itself. However, one has to take into account the symmetry relations
for $J^\mu(p_1,p_2,p_3,p_4)$. It is the
purpose of this appendix to show explicitly that knowledge of the
amplitude for the $2 \pi^- \pi^+ \pi^0$ mode is also sufficient to
determine all four matrix elements in the isospin limit.

For this purpose, we write the general decomposition of 
$J^\mu(p_1,p_2,p_3,p_4)$ as
\begin{eqnarray} 
J^\mu(p_1,p_2,p_3,p_4) &=& p_1^\mu B(p_1,p_2,p_3,p_4) +
p_2^\mu B(p_2,p_1,p_3,p_4) \nn
&+& p_3^\mu C(p_1,p_2,p_3,p_4) - p_4^\mu C(p_1,p_2,p_4,p_3)
\end{eqnarray} 
in terms of two invariant amplitudes that satisfy the constraints
\begin{eqnarray} 
B(p_1,p_2,p_3,p_4) & =& - B(p_1,p_2,p_4,p_3) \nn
C(p_1,p_2,p_3,p_4) & =& C(p_2,p_1,p_3,p_4) \label{eq:symmBC}
\end{eqnarray} 
due to  charge conjugation invariance and Bose symmetry. Gauge
invariance leads to a further relation between $B$ and $C$ but we 
do not need this relation here. 

The isospin relation (\ref{eq:k3}) can now be written as
\begin{eqnarray} 
\langle \pi^- \pi^- \pi^+ \pi^0 |V^\mu_{\rm cc}(0)
|0\rangle/\sqrt{2} &=& p_+^\mu D(p_+,p_1,p_2,p_0) +
p_1^\mu F(p_+,p_1,p_2,p_0) \nn
&+& p_2^\mu  F(p_+,p_2,p_1,p_0) - p_0^\mu G(p_+,p_1,p_2,p_0)
\end{eqnarray} 
with
\begin{eqnarray} 
D(p_+,p_1,p_2,p_0) &=& B(p_+,p_1,p_2,p_0)+B(p_+,p_2,p_1,p_0) \nn
F(p_+,p_1,p_2,p_0) &=& B(p_1,p_+,p_2,p_0)+C(p_+,p_2,p_1,p_0) \\
G(p_+,p_1,p_2,p_0) &=& C(p_+,p_1,p_0,p_2)+C(p_+,p_2,p_0,p_1)~. \no
\end{eqnarray}

With the symmetry relations (\ref{eq:symmBC}) one easily verifies 
\begin{eqnarray} 
2 B(p_1,p_2,p_3,p_4) & =& D(p_1,p_3,p_2,p_4)+ F(p_2,p_1,p_3,p_4)
-  F(p_3,p_1,p_2,p_4)\nn
2 C(p_1,p_2,p_3,p_4) & =& - D(p_3,p_2,p_1,p_4)+ F(p_1,p_3,p_2,p_4)
+ F(p_2,p_3,p_1,p_4) ~.
\end{eqnarray}
Therefore, the amplitude for the $2 \pi^0 \pi^+ \pi^-$ channel can be
obtained from the $2 \pi^- \pi^+ \pi^0$ amplitude. Consequently, 
knowledge of the $2 \pi^- \pi^+ \pi^0$ mode is enough to
determine the other three amplitudes in the isospin limit.

\setcounter{equation}{0}
\addtocounter{zaehler}{1}
\section{Vector and axial-vector mesons}
\label{app:VA}
Spin-1 mesons can be described either by the more conventional vector 
(or axial-vector) fields $V_\mu$ or by antisymmetric tensor fields
$V_{\mu\nu}$. For matching resonance exchange amplitudes with standard
CHPT amplitudes, the choice of fields is a matter of convenience. The
tensor field formalism has the advantage of producing immediately the
correct LECs of $O(p^4)$ \cite{gl84,egpr89}. On the other hand, single
resonance exchange that contributes first at $O(p^6)$ such as $\omega$
exchange is better described by vector fields \cite{eglpr89,epr90}.
The two descriptions are equivalent but the transformation from one
formalism to the other involves the introduction of explicit local
amplitudes. Those local terms may be avoided by employing the proper
formalism.

We recall first the usual normalization of a conventional massive 
vector field $V_\mu$ for a vector meson of mass $M$, with polarization
vector $\ve_\mu(p)$, and the associated propagator:
\begin{eqnarray}
\langle 0| V_\mu(0) | V,p\rangle &=& \varepsilon _\mu(p) \\
\langle 0|T\{ V_{\mu}(x),V_{\nu}(0)\}|0\rangle &=&
i \int \frac{d^4k e^{-ikx}}{(2\pi)^4(M^2-k^2-i\varepsilon)}
\left(g_{\mu\nu} - k_\mu k_\nu /M^2 \right) ~. 
\end{eqnarray} 

The same spin-1 particle can also be described by an antisymmetric
tensor field $V_{\mu\nu}$. The corresponding one-particle matrix
element and the propagator are given by \cite{egpr89}
\begin{eqnarray} 
\langle 0|V_{\mu\nu}(0)|V,p\rangle &=& i M^{-1} \{p_\mu \varepsilon_\nu(p) -
p_\nu \varepsilon_\mu(p)\} \\
\langle 0|T\{ V_{\mu\nu}(x),V_{\rho\sigma}(0)\}|0\rangle &=&
 i M^{-2} \int \frac{d^4k e^{-ikx}}{(2\pi)^4(M^2-k^2-i\varepsilon)}
\left[g_{\mu \rho} g_{\nu\sigma}(M^2-k^2) \right. \nn
&& + \left. g_{\mu\rho} k_\nu k_\sigma -
g_{\mu\sigma} k_\nu k_\rho - (\mu \leftrightarrow \nu)\right]~.
\label{eq:tprop1}
\end{eqnarray} 
In many cases, the tensor field propagator can be simplified. Whenever
the $\rho$ meson couples either directly to the (virtual)
photon or to two pions the transverse part of (\ref{eq:tprop1})
does not contribute \cite{gdpp00} and the
$\rho$ propagator may be replaced by 
\begin{eqnarray} 
\langle 0|T\{\rho_{\mu\nu}(x),\rho_{\rho\sigma}(0)\}|0\rangle &=&
 i \int \frac{d^4k e^{-ikx}}{(2\pi)^4(M^2-k^2-i\varepsilon)}
\left[g_{\mu \rho} g_{\nu\sigma} - g_{\mu \sigma} g_{\nu\rho}
\right] ~. \label{eq:tprop2}
\end{eqnarray} 
This happens to be the case for all diagrams considered
involving $\rho$ exchange.
The simplification does not apply for the $a_1$ propagator in the
diagrams of Fig.~\ref{fig:a1}.

\setcounter{equation}{0}
\addtocounter{zaehler}{1}
\section{Numerical input}
\label{app:num}
In this appendix we collect the numerical values of masses and
coupling constants that we have used for the calculation of
cross sections.

\vspace*{.5cm} 
\noindent
\Un{Chiral LECs}

\vspace*{.5cm} 
\begin{tabular}{lll}
$F_\pi$ = 0.0924 GeV &\hspace*{2cm} $\wt l_1$ = -2.0 $\times 10^{-3}$  
&\hspace*{2cm} $\wt l_2$ = 1.1 $\times 10^{-2}$ \\ 
$\wt l_3$ = -4.6 $\times 10^{-3}$ &
\hspace*{2cm} $\wt l_4$ = 2.8 $\times 10^{-2}$~ & 
\hspace*{2cm} $\wt l_6$ = -1.7 $\times 10^{-2}$
\end{tabular} 

\vspace*{.5cm} 
\noindent
\Un{Vector mesons}

\vspace*{.5cm} 
\begin{tabular}{lll}
$M_\rho$ = 0.775 GeV & \hspace*{2cm}  $F_V$ = 0.14 GeV &
\hspace*{1.5cm}  $G_V$ = 0.066 GeV  \\
$M_\omega$ = 0.783 GeV & \hspace*{2cm} $\Gamma_\omega$ = 0.00844 GeV 
& \hspace*{1.5cm} $g_{\omega\rho\pi}$ = 5.7
\end{tabular} 

\vspace*{.5cm} 
\noindent
\Un{Axial-vector meson}

\vspace*{.5cm} 
\begin{tabular}{ll}
$M_{a_1}$ = 1.23 GeV & \hspace*{2cm} $\Gamma_{a_1}$ = 0.5 GeV \\
$F_A = F_\pi$  & \hspace*{2cm} $c_2=c_3=c_4$ = 319
\end{tabular} 

\vspace*{.5cm} 
\noindent
\Un{Scalar mesons}

\vspace*{.5cm} 
\begin{tabular}{lll}
$M_{f_0}$ = 0.98 GeV & \hspace*{2cm} $\Gamma_{f_0}$ = 0.05 GeV & \\
$M_{\sigma}$ = 0.6 GeV & \hspace*{2cm} $\Gamma_{\sigma}$ = 0.6 GeV & \\
$c_d$ = 0.032 GeV & \hspace*{2cm} $c_m$ = 0.042 GeV & 
\hspace*{1.6cm} $\wt c_i$ = $c_i/\sqrt{3}~~~(i=d,m)$
\end{tabular} 


\end{document}